\documentclass[aip,graphicx,reprint]{revtex4-1}
\draft 

\usepackage{graphicx} 

\begin{document}
\title{Quantum simulations of charge-separation at a model donor-acceptor interface: role of delocalization and local packing.} 
\author{Allen C. Kelley}
\affiliation{Department of Chemistry, University of Houston, Houston TX 77204}
\author{Eric R. Bittner}
\email[]{bittner@uh.edu}
\homepage[]{k2.chem.uh.edu}
\affiliation{Department of Chemistry, University of Houston, Houston TX 77204}

\date{\today}

\begin{abstract}
We investigate the electronic dynamics of a model organic photovoltaic (OPV) system consisting of polyphenylene vinylene (PPV) oligomers and a [6,6]-phenyl C61-butyric acid methylester (PCBM) blend using a mixed molecular mechanics/quantum mechanics (MM/QM) approach. Using a heuristic model that connects energy gap fluctuations to the average electronic couplings and decoherence times, we provide and estimate of the state-to-state internal conversion rates within the manifold of the lowest few electronic excitations. We show that the electronic dynamics of the OPV are dramatically altered by varying the positions of the molecules simulated at the interface. The lowest few excited states of the model interface rapidly mix allowing low frequency C-C out of plain torsions to modulate the potential energy surface such that the system can sample both intermolecular charge-transfer and charge-separated electronic configurations on sub 100 fs time scales. Our simulations support an emerging picture of carrier generation in OPV systems in which interfacial electronic states can rapidly decay into charge-separated and current producing states via coupling to vibronic degrees of freedom. 
\end{abstract}


\maketitle 

\section{Introduction}

Advances in both materials and device fabrication have lead to the development 
of highly efficient organic polymer-based photovoltaic cell (OPV) in which the 
power conversion efficiency is in excess of 10-11\% under standard solar illumination~\cite{He:2012uq}
and efficiencies as high as 12\% in multi-junction OPVs.
This increase in power conversion efficiency indicates that mobile charge carriers can be efficiently generated
and collected in well-optimized devices;
however, the underlying photo-physical mechanism for converting highly-bound molecular (Frenkel) excitons into mobile 
and asymptotically free photocarriers 
remains elusive in spite of vigorous, multidisciplinary research activity. \cite{Tong2010,doi:10.1021/jz4010569,Gelinas:2013fk,Collini16012009,doi:10.1021/jp4071086,PhysRevLett.114.247003,Provencher:2014aa,Engel:2007aa,Beljonne:2005,tamura:021103,Grancini:2013uq,C3FD20142B,tamura:107402,Sheng2012,PhysRevB.71.045203,ja4093874,doi:10.1021/jp104111h} 

The generic photophysical pathway that underlay the generation of mobile charge carriers in the OPV begins with an exciton being formed inside the system. The exciton dissociates at the interface into a charge transfer state with a small electron/hole separation. The charge transfer state is bound by Coulombic attraction to the interface, impeding the ability of the particles to migrate away from the interface to form free charge carriers. Ultrafast spectroscopic measurements on OPV systems have reported that charged photoexcitations are generated on $\leq$ 100-fs \cite{Tong2010,Grancini:2013uq,Sheng2012,Sariciftci:1994kx,Banerji:2010vn,Banerji:2013ej,Jailaubekov:2013fk} time-scales, despite the strong Coulombic attraction due to the low dielectric constant prevalent in OPV's. Experiments by Gelinas $et$ $al$., in which Stark-effect signatures in transient absorption spectra were analysed to probe the local electric field as charge separation proceeds, indicate that electrons and holes separate by as much as 40\AA \hspace{1pt} over the first 100 fs and evolve further on pico-second time scales to produce unbound and hence freely mobile charge pairs.\cite{Gelinas:2013fk}

Bittner and Silva recently presented a fully quantum dynamical model of photo-induced charge-fission at a polymeric type-II heterojunction interface.\cite{Bittner:2014aa} This model supposes that the energy level fluctuations due to bulk atomic motions create the resonant conditions for coherent separation of electrons and holes via long-range tunnelling. Simulations based upon lattice models reveal that such fluctuations lead to strong quantum mechanical coupling between excitonic states produced near the interface and unbound electron/hole scattering states resulting in a strong enhancement of the decay rate of photoexcitations into unbound polaronic states.

A microscopic model of the interface is required to understand the mechanisms that promote ultra-fast charge separation. This requires knowing the morphology and the finest details of the interface and its electronic structure. Currently {\em ab inito} methods provide a good tool for exploring the formation of charge transfer states in donor/acceptor pairs, however, it is ill suited for the task of simulating the formation of charge separated states, because it is to computationally cumbersome to expand the interface from a single donor/acceptor pair into the size needed to allow charges separation to occur.


In this paper, we take a molecular mechanics/quantum mechanics (MM/QM) approach to study Poly(p-phenylene vinylene) (PPV)/ Phenyl-C61-butyric acid methyl ester (PCBM) heterojunctions. Polymer microstructural probes have revealed general relationships between disorder, aggregation and electronic properties in polymeric semiconductors.\cite{Noriega:2013uq} The distribution of torsion angles for the PPV molecules at the interface are larger than in the bulk, adding to the structural disorder of the PPV molecules closest to the interface. Moreover, aggregation (ordering) can be perturbed by the blend-ratio and composition of the donor/acceptor polymers. \cite{Noriega:2013uq} We explore the effect that the positioning of the molecules at the interface has on the electronic properties and estimate the state-state transition rate constants of the systems. 

\section{Methods}

Our simulations employ a modified version of the TINKER molecular dynamics (MD) package\cite{ponder:2004} in which the MM3 \cite{allinger:868} intra-molecular bonding parameters are allowed to vary with the local $\pi$-electronic density 
as described by a Parisier-Parr-Pople (PPP) semi-empirical Hamiltonian
\cite{pople:1375,pariser:767}. 
Similar approaches have been described by Rossky\cite{Rossky2000} and Warshal\cite{Warshel1995} to include electronic dynamics into an otherwise classical force field description.


At each time-step of the simulation, we compute the Hartree-Fock (HF) ground state for the $\pi$ system and use configuration interaction (singles) (CI-S) to describe the lowest few $\pi \rightarrow \pi^*$ excitations. 
Intermolecular interactions within the active space are introduced via non-bonding Coulombic coupling terms and static dispersion interactions contained within the MM3 forcefield.  All electronic excitations are confined to the $\pi$-active orbitals. We used a total of 10 occupied and 10 unoccupied Hartree Fock molecular orbitals to construct the electron/hole configurations for the CI calculations. The excited state bond charge density matrix is constructed by assuming that electron densities are added to the virtual orbitals and hole densities are subtracted from the Hartree Fock ground state. The equilibrium bond orders are modified by

\begin{eqnarray}
BO_{i}=I_{BO}(i) +T_{BO}(1-P(i))
\end{eqnarray}
where $I_{BO}$ and $T_{BO}$ are the initial equilibrium bond order and the rate of bond length increase with bond order decrease found in the MM3 parameter set, $P$ is the excited state bond order found in the excited state bond charge density matrix. Modifying the equilibrium bond order allows the bond lengths in the MM calculation to change in response to local changes in bond order due to the migration of charge.

During the equilibration steps we assume the system to be in its electronic ground state, after which we excite the system to the first CIS excited state and allow the system to respond to the change in the electronic density within the adiabatic Born-Oppenheimer approximation. It is important to note that the excited state we prepare is not the state which carries the most oscillator strength to the ground nor do we account for non-adiabatic surface hopping-type transitions in our approach.\cite{tully:1061,tully:562,Granucci:2007} The dynamics simulations shown reflect the longer-time fate of the lowest-lying excited state populations and sample possible configurations that can be accessed by the system. The combination of a classical MD forcefield with a semi-empirical description of a selected few molecules within the system seems to be a suitable compromise between a fully {\em ab initio} approach which would be limited to only a few molecules and short simulation times and a fully classical MD description which would neglect any transient changes in the local electronic density.\cite{Jailaubekov:2013fk} 
  

We specifically chose three separate clusters of molecules to represent model bulk-hetrojunctions in order to study how varying the blend and positioning of the molecules affect the penetration of extended intra molecular electronic states into the bulk region. In each case study we selected a subgroup of PCBM molecules and PPV chains, treating the $\pi$-electrons in these units explicitly while the remaining molecules in the simulations are treated using the purely classical MM3 force-field. The number of $\pi$ active PCBM molecules vary between each simulation, allowing each system to have a different blend ratio inside the $\pi$ active system. The placement of PPV molecules vary in two of the simulations, changing the number of PPV molecules in direct contact with PCBM molecules, fundamentally changing the hetrojunction.   

\begin{figure}
\centering
\includegraphics[height=6.5cm]{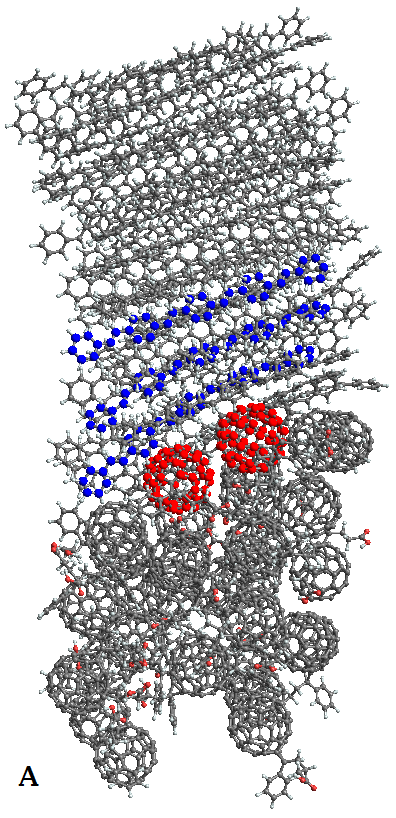}
\includegraphics[height=6.5cm]{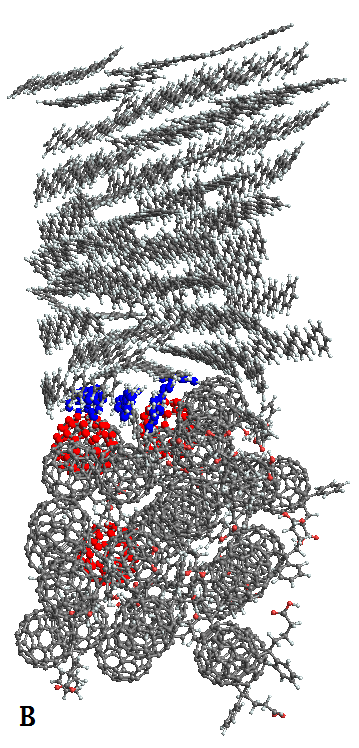}
\includegraphics[height=5.5cm]{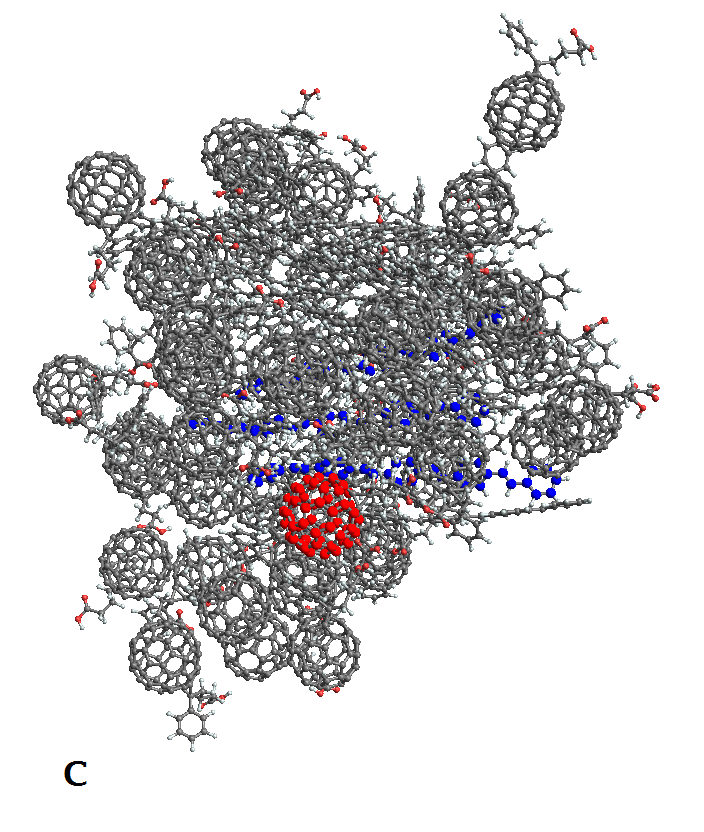}
\caption{Snapshot of the three MD simulation cells: A, B and C. The red and blue highlighted molecules denote the $\pi$-active units in our system. }
\label{fig1}
\end{figure}

Figure~\ref{fig1} shows 
representative configurations  
of the three cases studied. In each, the red and blue coloured spheres represent atoms included in the quantum-chemical description. Each corresponds to a periodic simulation cell in the $xy$ plane following equilibration at 100K and 1 atm pressure. In Case A, we selected 2 interfacial PCBMs and 3 nearby $\pi$-active PPV oligomers that penetrate into the bulk polymer region, including a total of 230 carbon $2p_z$ orbitals. In Case B, we selected 3 PCBM and 3 nearby PPV oligomers expanding the $\pi$ active molecules that form the inter-facial hetrojunction, including a total of 288 $2p_z$ orbitals. The system has the same boundary conditions as case A and is set up such that the main difference is in the placement of the PPV oligomers. In simulation C we selected 1 PCBM and 3 PPV oligomers consisting of 172 carbon $2p_Z$ that penetrate into the bulk using and employ periodic boundary conditions in $xyz$. This simulation was set up to be very similar to case A, only adding a single PCBM molecule to the inter-facial region. 
The case studies are representative of typical interfacial configurations and in 
no way are a comprehensive sampling. 




\section{Results}

\begin{figure}
\hspace{6pt}\includegraphics[height=4.0cm]{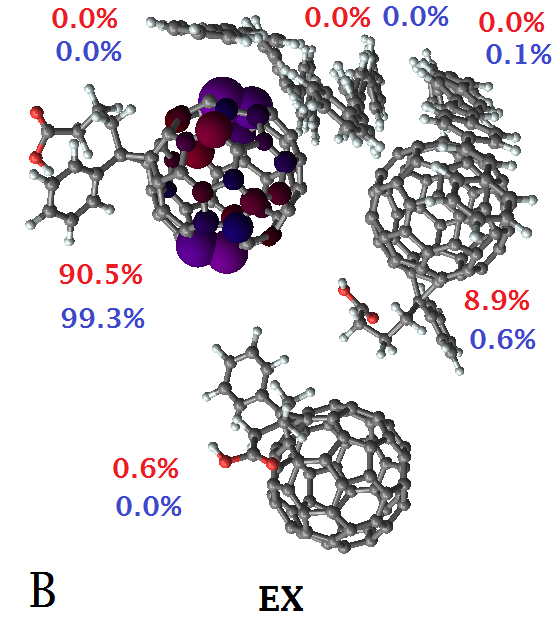}\hspace{35pt}
\includegraphics[height=4.0cm]{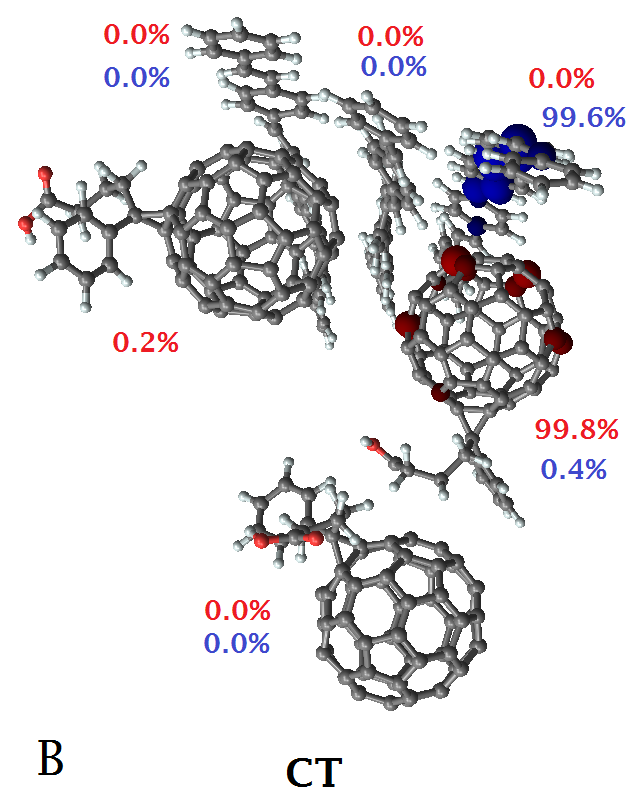}
\\
\includegraphics[height=3.5cm]{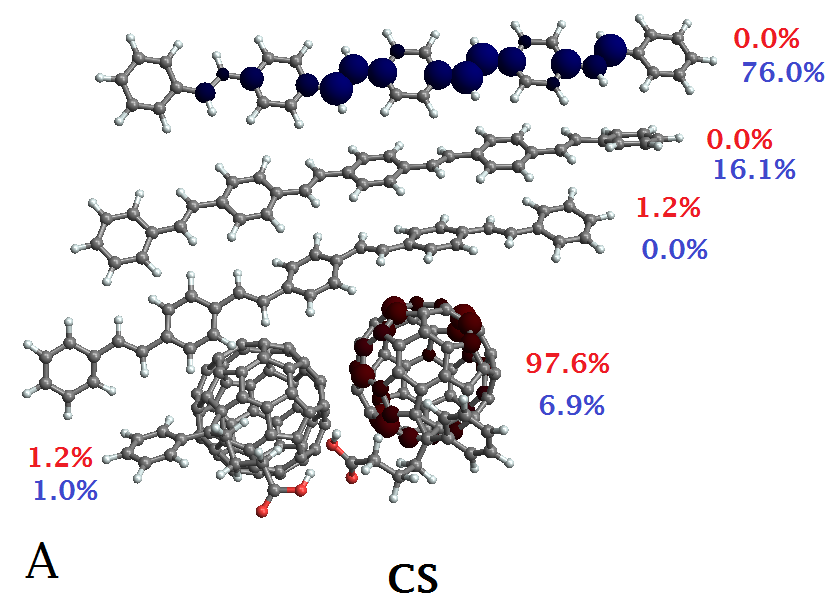}
\includegraphics[height=3.5cm]{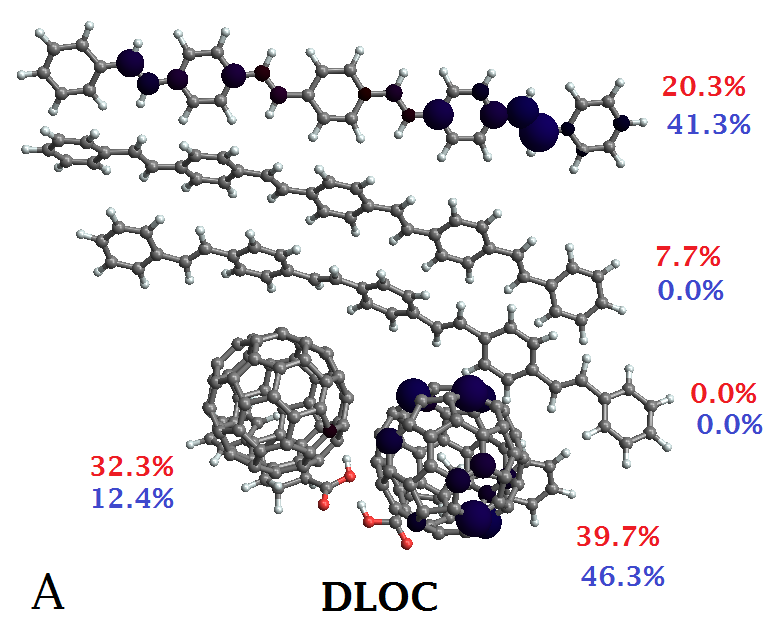}
\caption{The dominant adiabatic states from simulation A (Bottom) and B (Top) as shown in Figure~\ref{fig1} The red and blue numbers denote the electron/hole density as a percent on the indicated molecule. The states correspond to the $x$ axis in Figure~\ref{fig3}. The four snapshots shown represent typical states for our systems. The top pictures from left to right present an exciton located on the PCBM molecule and a charge transfer state. The bottom pictures from left to right show a charge separated state and a partly de-localized state.}
\label{fig2}
\end{figure}

\begin{figure}
\centering
\includegraphics[height=4cm]{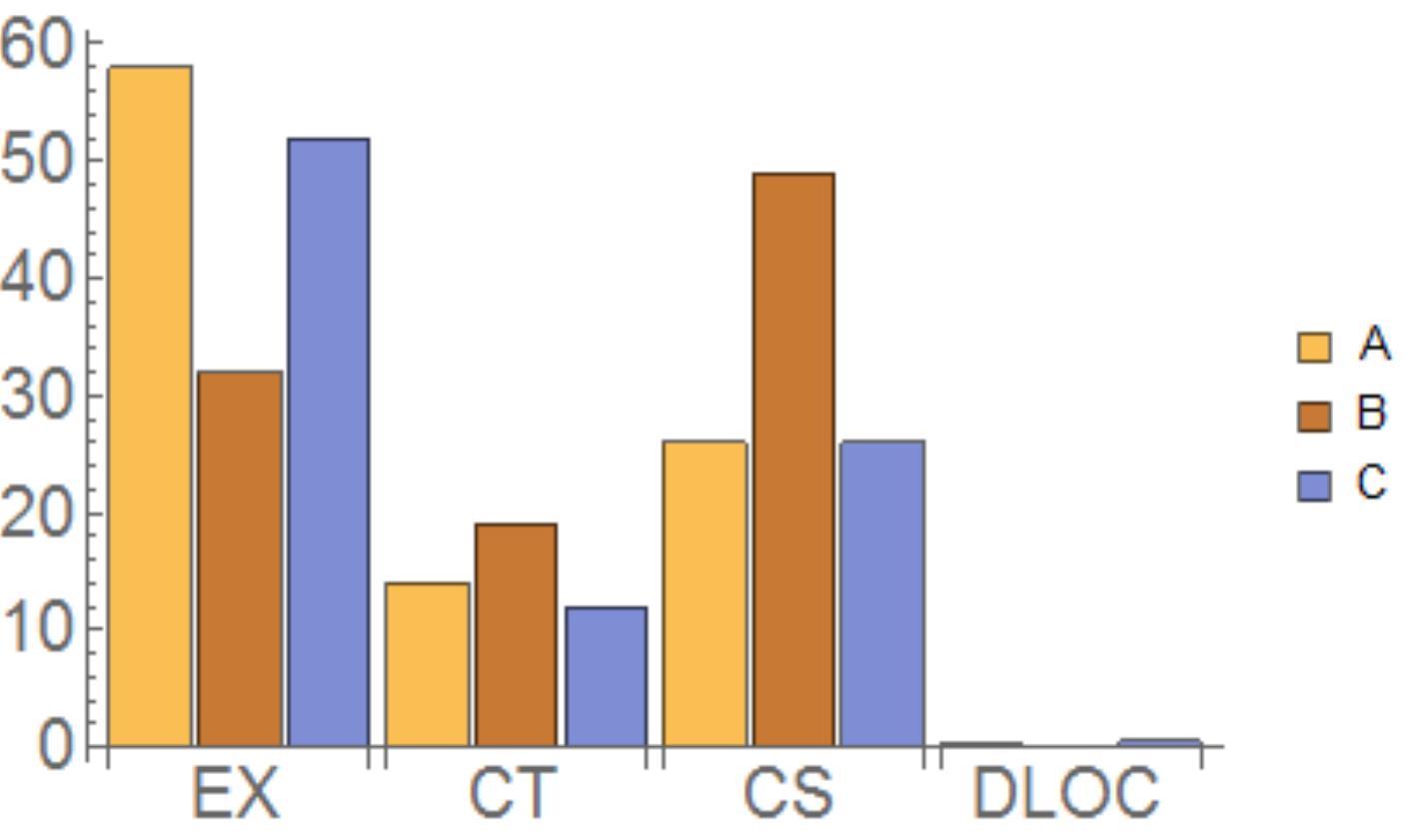}
\caption{Histogram plot of the populations of adiabatic states in system A, B and C taken every fs. Each cluster represents a different classification of the adiabatic states as shown in Figure ~\ref{fig2}.}\label{fig3}
\end{figure}

Over the course of eight 10 ps simulations, the lowest lying CIS excitation samples a variety of adiabatic states ranging from localized excitons to charge-separated, charge-transfer and de-localized configurations. 
Figure~\ref{fig2} shows the configurations of four distinct adiabatic states, where the probability of the system to be found in a specific state can be seen in Figure~\ref{fig3}. We categorize the states into four types, (EX) represents the exciton states, characterized as having >50\% of the electron/hole density on a single molecule. The exciton state populates $\approx$ 58\% of the states for simulation A, $\approx$ 32\% for simulation B and $\approx$ 52\% for simulation C. (CT) represents the charge-transfer states, characterized as having >50\% of the electron/hole density occupying adjacent molecules. The charge-transfer state populates $\approx$ 14\% of the states for simulation A, $\approx$ 19\% for simulation B and $\approx$ 12\% for simulation C. (CS) represents the charge separated states, characterized as having >50\% of the electron/hole density occupying a PCBM and a PPV separated by a single molecule. The charge separated state populates $\approx$ 26\% for simulation A, $\approx$ 49\% for simulation B and $\approx$ 26\% for simulation C. (DLOC) represents the de-localized states, characterized as having <50\% of the electron/hole density on a single PPV or PCBM molecule.

\begin{figure}
\centering
\includegraphics[height=9cm]{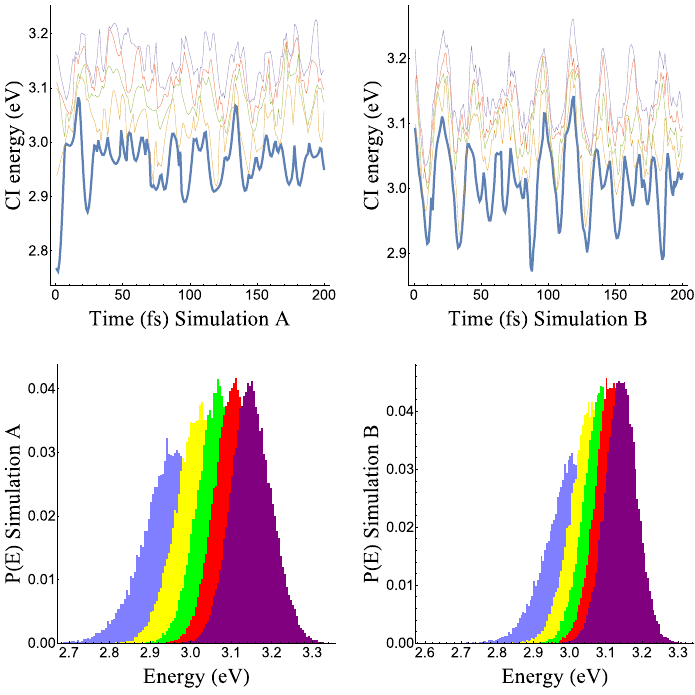}
\caption{(Top) Single CI (SCI) energies for simulation A and B for the first 200 fs following excitation to the lowest SCI state. (Bottom) Histogram distributions of the 5 lowest excitation energy levels over a 200 ps simulation. Throughout the simulation, the lowest lying state 
(in blue)   
remained populated and varied in character from excitonic to charge-separated. }\label{fig4}
\end{figure}

The energies of the lowest CIS state following excitation at $t = 0$ fs for simulation A and B are shown in Figure~\ref{fig4} (Top). After excitation at $t = 0$ there is very little energetic relaxation in all of the systems simulated. The simulations appear to cycle through many adiabatic states in a short period of time leaving the impression of a weak electron-phonon coupling. This can be rationalized as the electron/hole density often de-localize over multiple molecules and many conjugated C-C bonds. Another striking affect of the systems is the large number of avoided crossings that occur between the lowest lying states. There is also a 20 fs oscillation in the CI energies, driving the systems excited states into many regions of strong coupling. The 20 fs oscillation also appears in the autocorrelation and and the Fourier transform of the gap energies, contributing the C=C bond stretching modes around $\approx$ 1600 cm$^{-1}$. The oscillation is contributed to small thermally activated fluctuations within the simulation, showing that even at 100 K the thermal fluctuations possess sufficient energy to bring these states into regions of strong electronic coupling. 

In Figure~\ref{fig4} (Bottom) we show a histogram of the 5 lowest CIS energies accumulated over 40 ps of simulation time following promotion to the lowest CIS state. The energies take a Gaussian distribution around the mean and contain large regions of overlap. The range and mean CI energy for the first two excited states (blue and yellow) in simulation A are 2.5-3.17 eV and 2.82-3.22 eV and 2.95 eV and 3.02 eV and in simulation B 2.69-3.19 eV and 3.00 eV and 3.05 eV. A mean band-gap of 0.070 eV and 0.050 eV between the first and second excited state allows a mechanism for the formation of a continuum of excited states that can easily be brought into strong electronic coupling by small fluctuations in the CI energy of the system. The small average band gap and rapid oscillatory nature of the CI energies facilitate the systems ability to rapidly sample a great many different electronic configurations over the course of the simulation.

We next consider the origins of the energy fluctuations evidenced in Figure~\ref{fig4} (Top). While we only show two 200 fs segments of eight 10 ps simulations over this period, one can see that the CI energies are modulated and cover a small range. The autocorrelation plots of the band gap energies, shown in Figure 5, show that the correlation times for the three simulations are very short $\approx$ 8 fs meaning that the system changes rapidly enough that the oscillations observed are independent of one another. By taking the Fourier transform of the gap the IR active modes that contribute to the modulation of the CI energies inside of the systems are found as shown in Figure 5. The modulation of the CI energy appears to be heavily dependent upon the torsion, C=C, and C-H stretching modes. In each of the plots three distinct regions can be seen, the low frequency torsional modes occur between 200 and 500 cm$^{-1}$, the C=C stretching modes occur between 1300 and 1800 cm$^{-1}$ and the C-H stretching modes occur between 2800 and 3300 cm$^{-1}$. We conclude that small-scale vibronic fluctuations in the molecular structures and orientations produce significant energetic overlap between different adiabatic states to drive the system from purely excitonic to purely charge-transfer on a rapid time-scale. This is evidenced in the progression of the CI energies, as small fluctuations in these modes can easily bring the excited states into strong coupling regimes. 

\begin{figure*}
\centering
\includegraphics[width=0.6\columnwidth]{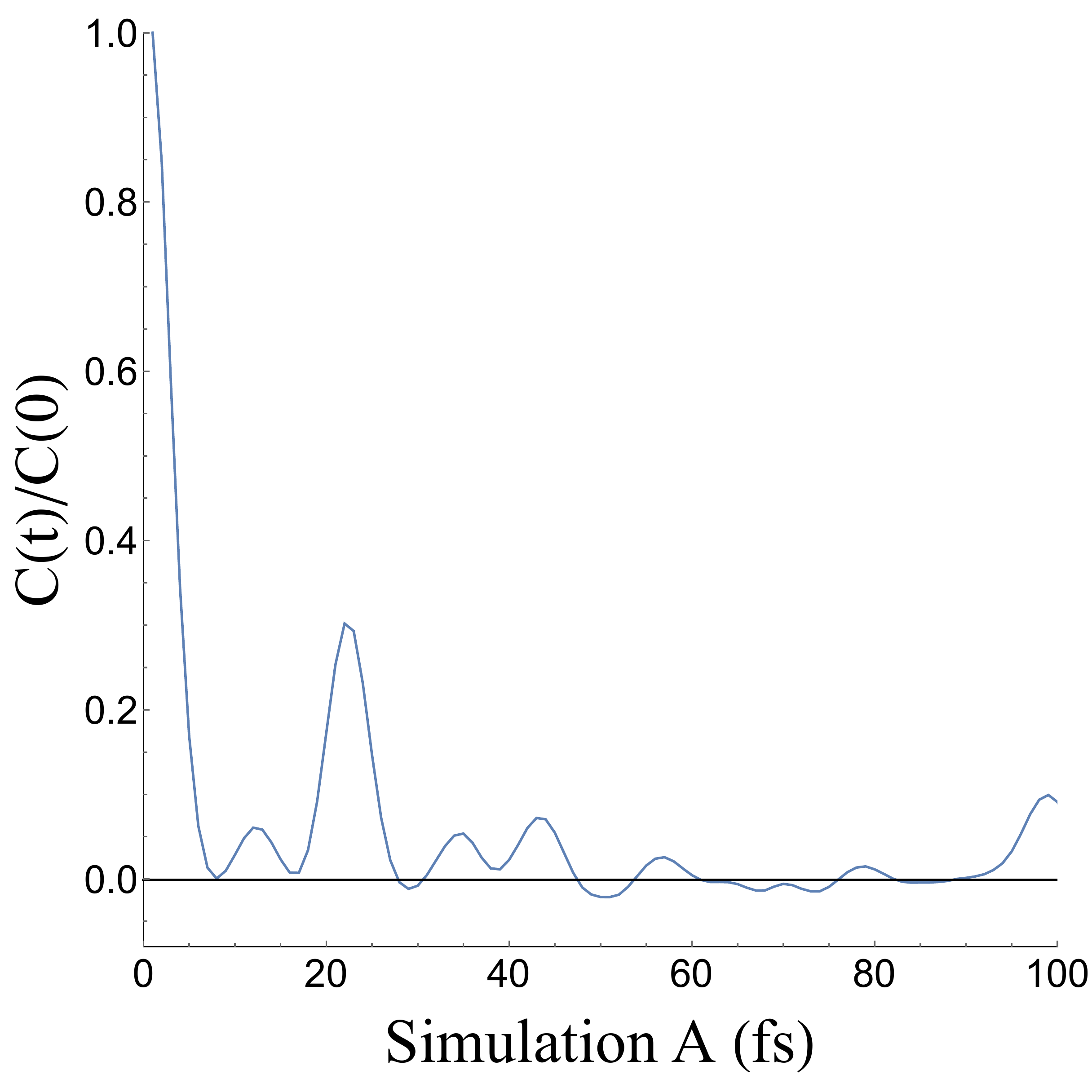}
\includegraphics[width=0.6\columnwidth]{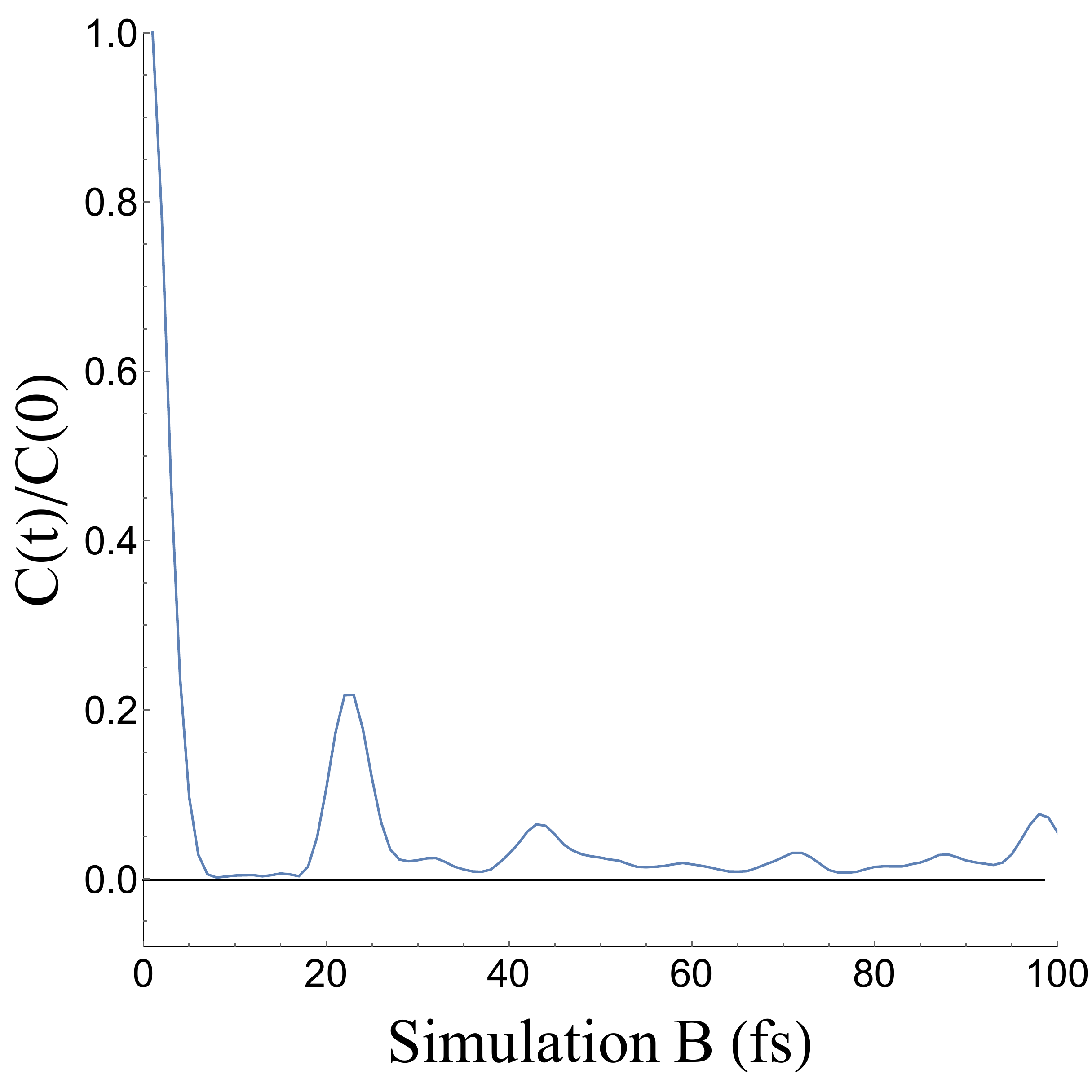}
\includegraphics[width=0.6\columnwidth]{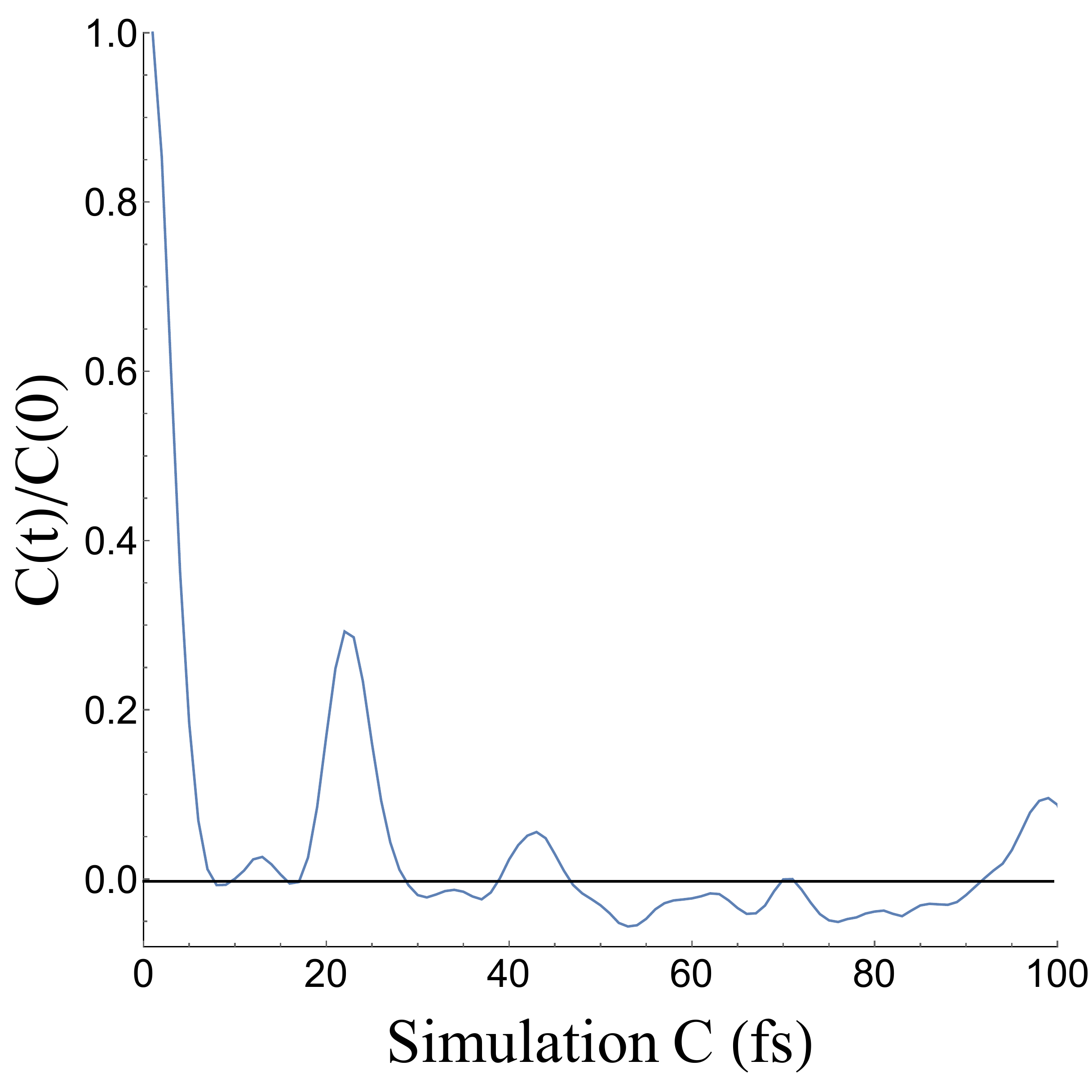}\\
\includegraphics[width=0.6\columnwidth]{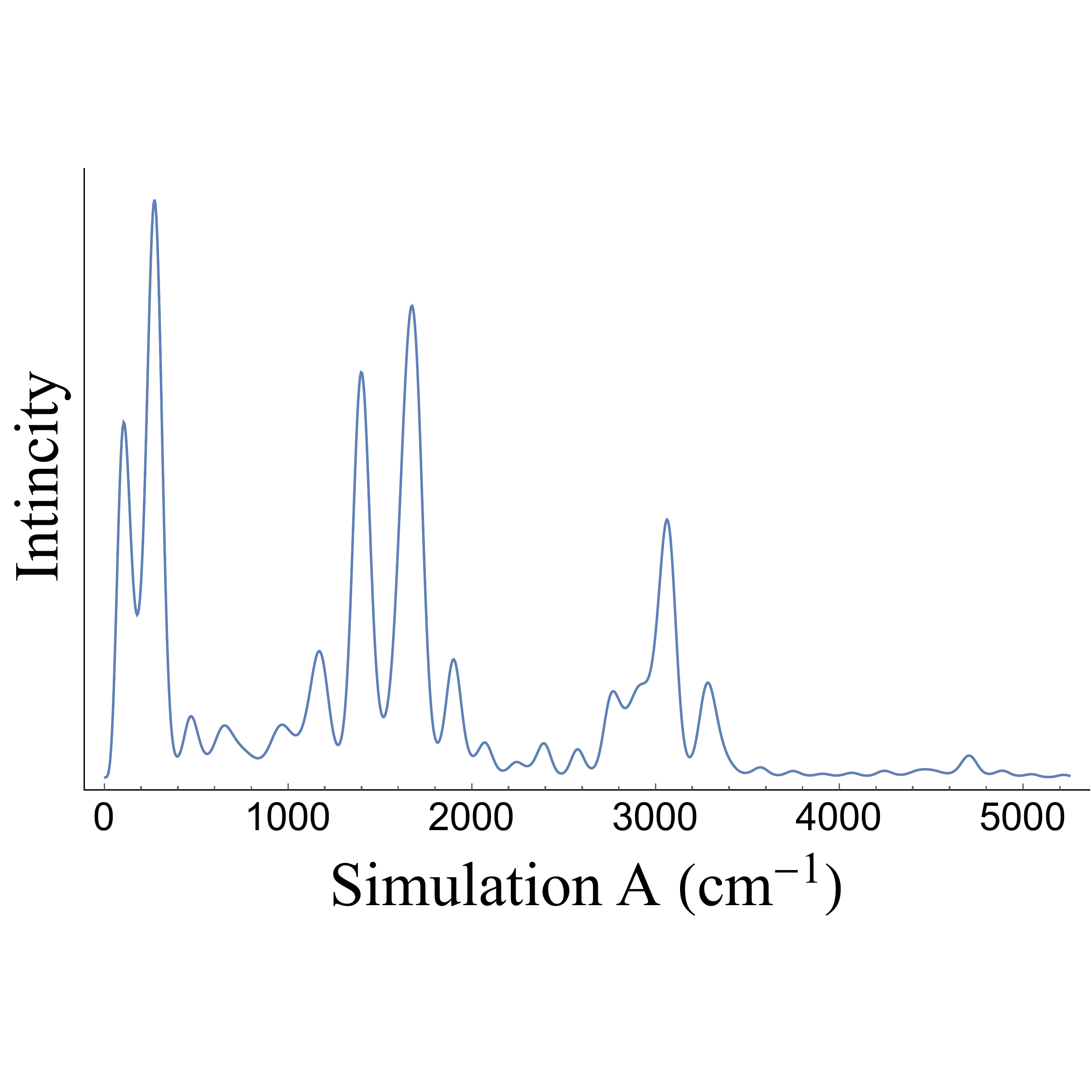}
\includegraphics[width=0.6\columnwidth]{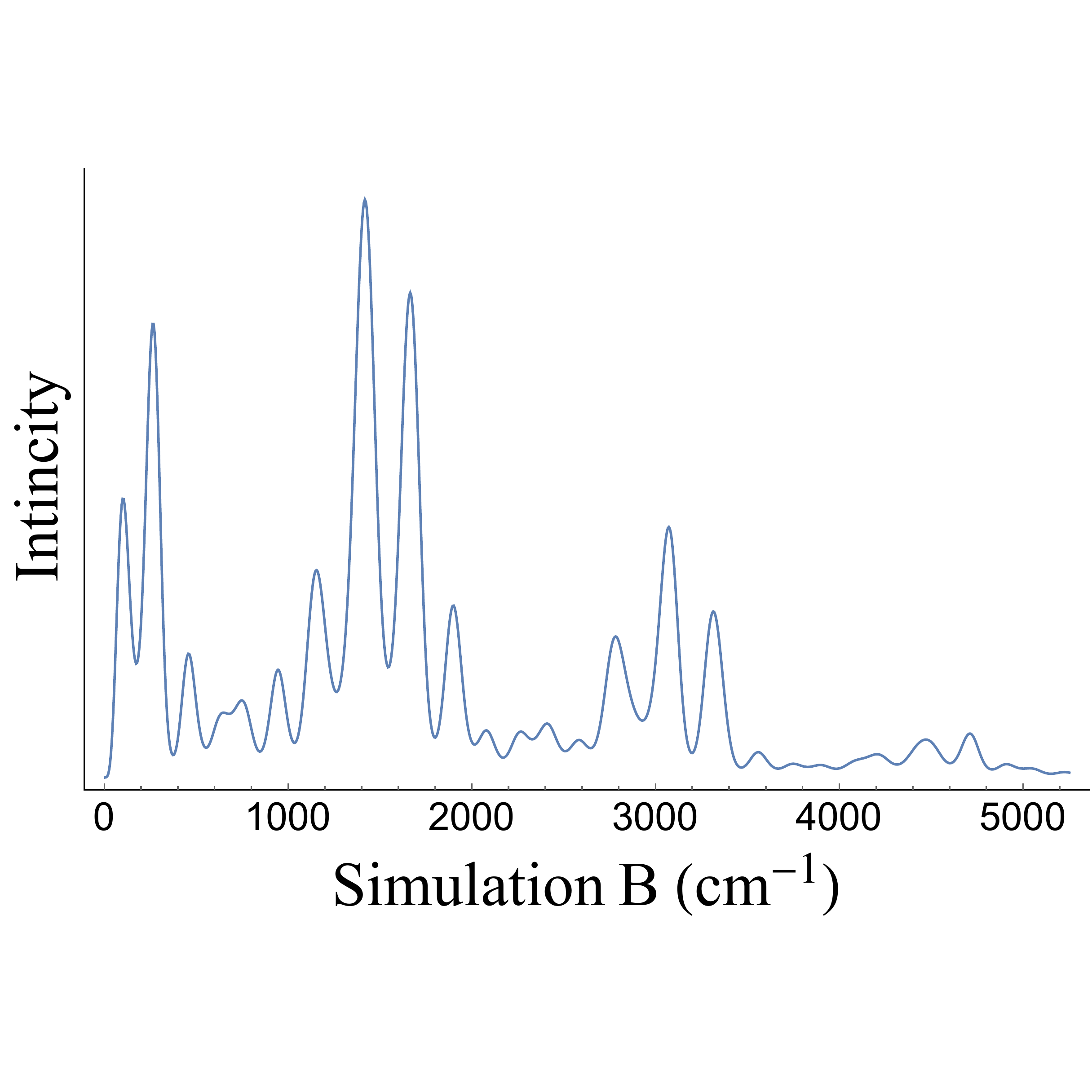}
\includegraphics[width=0.6\columnwidth]{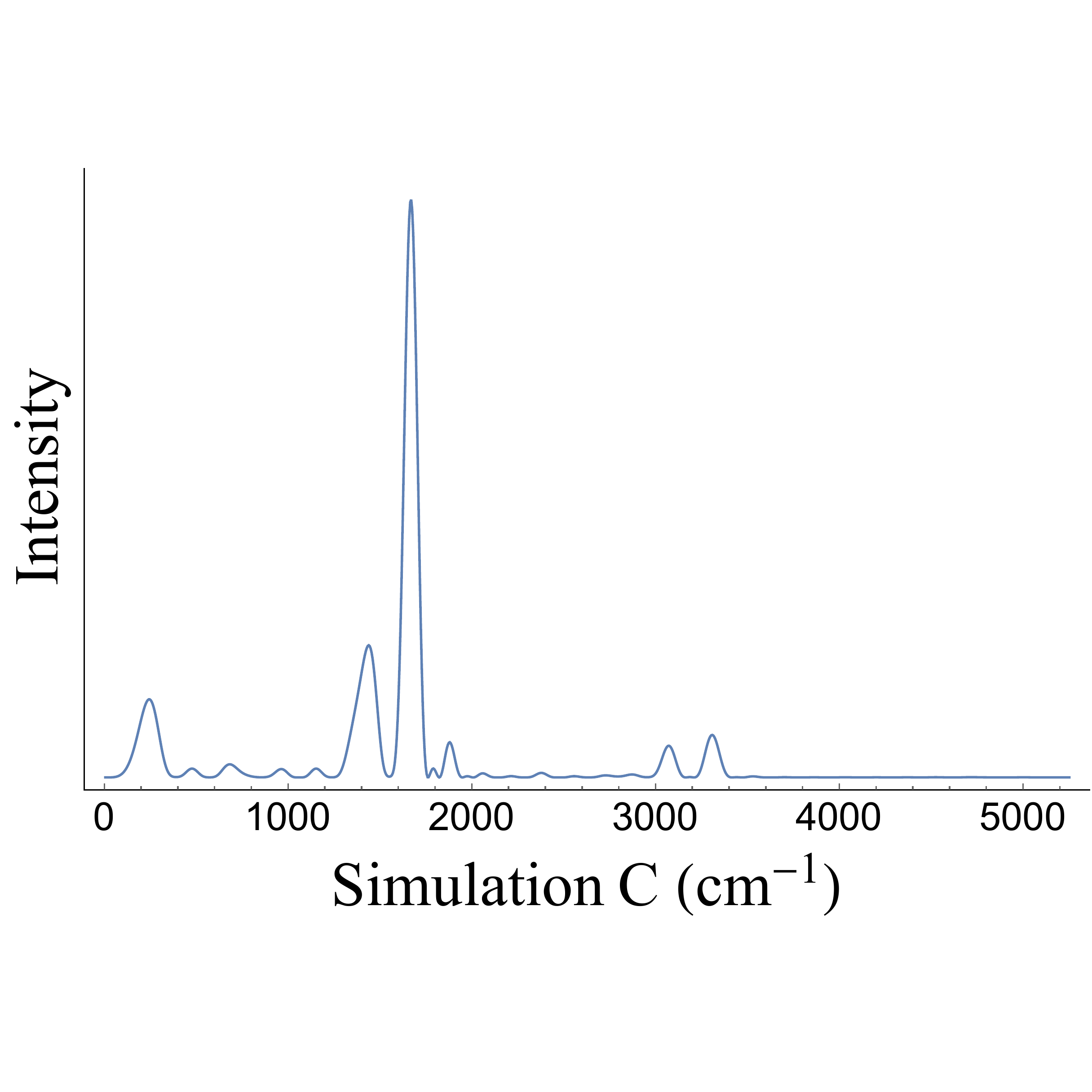}\\
\caption{(Top) Autocorrelation of the energy gap between first and second excited states
for simulations A, B and C. 
The correlation time for all three simulations are $\approx$ 8 fs. 
(Bottom) These plots show the active IR active modes that contribute to the modulation in the CI energies of the systems. Three distinct regions are visible in each plot. The low frequency torsional modes occur between 200 and 500 cm$^{-1}$, the C=C stretching modes occur between 1300 and 1800 cm$^{-1}$ and the C-H stretching modes occur between 2800 and 3300 cm$^{-1}$}
\end{figure*}

\subsection{Estimating state to state rates}

We can estimate the state $\rightarrow$ state rates using the model by Bittner\cite{Bittner:911,Bittner:2014aa}. Consider a two state system with coupling $\lambda$ in which the energy gap $\Delta$(t) fluctuates in time around its average $\bar{\Delta}$. In a two state basis the Hamiltonian can be written as
\begin{eqnarray}
H=\frac{\Delta(t)}{2}\hat{\sigma_z}+\lambda\hat{\sigma_x} 
\end{eqnarray}
where $\hat{\sigma_k}$ are Pauli matrices. Note that Eq. 1 transformed such that fluctuations are in the off-diagonal coupling, becomes
\begin{eqnarray}
H=\frac{\Delta_0}{2}\hat{\sigma_z}+\delta V(t)\hat{\sigma_x} 
\end{eqnarray}
where $\Delta_0 =\bar{\Delta}+\lambda$ and $\delta\bar{V}(t)=0$. The fluctuations in the electronic energy levels are attributed to thermal and bond-vibrational motions of polymer chains which can be related to the spectral density, $S(\omega)$ via 
\begin{eqnarray}
\bar{V}^2=\delta\bar{V}^2 (t)=\int_{-\infty}^{+\infty}\frac{d\omega}{2\pi}S(\omega). 
\end{eqnarray}
Averaging over the environmental noise, we can write the average energy gap as $\hbar\bar{\Omega}$=$\sqrt[]{\Delta^2_0 + \bar{V}^2}$ with eigenstates 
\begin{eqnarray}
\centering
\vert\Psi+\rangle = \cos \theta\vert0\rangle + \sin \theta\vert1\rangle
\end{eqnarray}
\begin{eqnarray}
\centering
\vert\Psi-\rangle = -\sin \theta\vert0\rangle + \cos \theta\vert1\rangle
\end{eqnarray}
where $\tan2\theta = \vert\bar{V}\vert/\Delta_0$ defines the mixing angle between original kets. Consequently, by analyzing energy gap fluctuations, we can obtain an estimate of both the coupling between states as well as transition rates. To estimate the average transition rate between states the equations of motion for the reduced density matrix for a two level system coupled to a dissipative environment are used.\\

$\hspace{33pt}\dot{\rho}_{11} =-\frac{i}{\hbar}V(\rho_{21}-\rho_{12})-\frac{1}{\tau_1}\rho_{11}$\\ 
$_{}\hspace{41pt}\dot{\rho}_{22} =\frac{i}{\hbar}V(\rho_{21}-\rho_{12})-\frac{1}{\tau_2}\rho_{22}$
\begin{eqnarray}
\dot{\rho}_{12} =-\frac{i}{\hbar}V(\rho_{22}-\rho_{11})-\frac{1}{T_d}\rho_{12}-\frac{\Delta_0}{i\hbar}\rho_{12} 
\end{eqnarray}
$_{}\hspace{42pt}\dot{\rho}_{21} =\frac{i}{\hbar}V(\rho_{22}-\rho_{11})-\frac{1}{T_d}\rho_{21}+\frac{\Delta_0}{i\hbar}\rho_{21}$\\

$\tau_1$ and $\tau_2$ have been introduced as the lifetimes of each state and $T_d$ is the decoherence time for the quantum superposition. The decoherence time can be related to the spectral density via $T^{-1}_{d}$ = $\bar{V}/\hbar$. Taking $T_d$ to be short compared to the lifetimes of each state, we can write the population of the initial states as

\begin{eqnarray}
\rho_{11}(t)= \exp[-\left(\frac{1}{\tau_1}-k\right)]
\end{eqnarray}
where $k$ is the average state to state transition rate. If we integrate over all time we obtain an equation of the form 

\begin{eqnarray}
\int_0^\infty \rho_{11}(t)dt = \left(\frac{1}{\tau_1}+k\right)^{-1}
\end{eqnarray}
suggesting a form for the exact solution of Eqs.6. Taking the Laplace transform of the equations of motion (Eqs.6) and assuming that our initial population is in state 1 ($\rho_{11}(0)=1$) the equations of motion become a series of algebraic equations\\

$_{}\hspace{38pt}-1=-\frac{i}{\hbar}V(\rho_{21}-\rho_{12})-\frac{1}{\tau_1}\rho_{11}$\\

$_{}\hspace{47pt}0=\frac{i}{\hbar}V(\rho_{21}-\rho_{12})-\frac{1}{\tau_2}\rho_{22}$

\begin{eqnarray}
\hspace{20pt}0=-\frac{i}{\hbar}V(\rho_{22}-\rho_{11})-\frac{1}{T_d}\rho_{12}-\frac{\Delta_0}{i\hbar}\rho_{12}
\end{eqnarray}
$_{}\hspace{56pt}0=\frac{i}{\hbar}V(\rho_{22}-\rho_{11})-\frac{1}{T_d}\rho_{21}+\frac{\Delta_0}{i\hbar}\rho_{21}\newline$\\
which after a bit of algebra gives a rate constant of the form.

\begin{eqnarray}
k = 2\frac{\bar{V}^2}{\hbar^2}\frac{T_d}{(T_d\Delta_0/\hbar)^2+1} 
\end{eqnarray}

According to the model outlined above the mean ($\Delta_0$) and variance (\={V}) of the energy gap distributions shown in Fig. 4 can be used as input to estimate the state to state transition rate for a two level system. We take $T_{\tiny{d}}^{-1}$ $\approx \bar{V}/\hbar$ as an estimate of the decoherence time and we introduce $\tau$ as the natural lifetimes of each state. The results are shown in Table 1. The estimated transition rates are consistent with the observations that the systems rapidly sample a wide number of possible configurations over the course of the molecular dynamics simulation. On average, the state to state couplings of 56 meV for simulation A and 48 meV for simulation B are comparable to the average energy gaps between the lowest excited states. The strong electronic coupling allows for rapid transitions; however, larger couplings also imply shorter electronic decoherence times, effectively quenching the ability of charges to separate by tunnelling. 

\begin{table}[h]
\small
  \caption{\ Estimated interstate transition rates and vibronic couplings for simulation A, B and C.
}
  \label{tbl:example}
  \begin{tabular*}{0.5\textwidth}{@{\extracolsep{\fill}}lllll}
    \hline
    Transition & $\Delta_o$ (eV) & $\langle V\rangle$ (eV) & $T_d$ (fs) &  $k^{-1}$ (fs) \\
    \hline
    \\
    A\\
    1$\rightarrow$2 & 0.070 & 0.050 & 13.16 & 19.5 \\
    1$\rightarrow$3 & 0.12 & 0.056 & 11.75 & 32.9 \\
    1$\rightarrow$4 & 0.16 & 0.063 & 10.45 & 38.9 \\
    \\
    B\\
    1$\rightarrow$2 & 0.050 & 0.042 & 15.67 & 18.9\\
    1$\rightarrow$3 & 0.09 & 0.049 & 13.43 & 29.3\\
    1$\rightarrow$4 & 0.12 & 0.052 & 12.66 & 40.0\\
    \\
    C\\
    1$\rightarrow$2 & 0.11 & 0.057 & 11.41 & 21.8\\
    1$\rightarrow$3 & 0.18 & 0.076 &  8.66 & 28.6\\
    1$\rightarrow$4 & 0.26 & 0.076 &  8.70 & 50.0\\
    \\
    \hline
  \end{tabular*}
  \label{table1}
\end{table}


\section{Conclusions}

We present here the results of hybrid QM/MM simulations of the excited states of model PPV/PCBM heterojunction interfaces. Our results indicate that varying the blend ratio and placement of the molecules comprising the heterojunction greatly affect the distribution of states yet have little affect upon the rate constants of the system. We also propose that thermal noise can rapidly change the character of the lowest lying excited state from purely excitonic to charge separated on a time scale of sub 100 fs.   

Simulations A and C have a very similar placement of molecules, only differing in that simulation A adds a PCBM molecule to the heterojunction. The addition of the PCBM only slightly changes the distribution of states as seen in Figure 4. The exciton states continue to be the most favored state inside the system, even slightly increasing in probability, while the delocalized states slightly decrease in probability. Simulation B completely changes the heterojunction, placing three PPV and two PCBM molecules at the interface as seen in Figure 1. The states generated by simulation B are radically different from those seen in simulation A and C as seen in Figure 4. The probability of finding the system in the exciton state is dramatically reduced while the charge separated state becomes predominant. This result is quite interesting as it highlights that the complexity of simulating heterojunctions resides not only in the size of the system but on how the donor/accepter interface is chosen. 

All three systems start with an exciton localized on the PCBM and dissociating into a charge transfer state with the hole (or electron) delocalized over multiple polymer units before localizing to form charge separated states. There are a wide range of electronic states tightly clustered within a small energy band, allowing small changes in local bond lengths to have a dramatic role in modulating the electronic couplings between excited states. We speculate that the dramatic shift in population seen in simulation B can be caused by disorder in the PPV molecules reducing the band gap by 20 meV. The PPV molecules comprising the interface region undergo large distortions in the C-C torsion angles allowing the molecules to cycle through a larger range of configurations inside of a short time interval. The presence of more $\pi$ active PPV molecules at the interface also appear to lead to more avoided crossing regions and the ability of the system to more efficiently dissociate excitons into charge transfer and charge separated states to a distance to where their Coulombic attraction is comparable to the thermal energy. While the finite size of our system prevents further dissociation of the charges, the results are suggestive that such interstate crossing events driven by bond-fluctuations can efficiently separate the charges. The results presented here corroborate recent ultra-fast experimental evidence suggesting that free polarons can form on ultra fast time scales (sub 100 fs) and that thermally activated low frequency torsional modes are key in effective electron hole separation in PPV/PCBM heterojunctions.

\section*{Acknowledgements}
The work at the University of Houston was funded in part by the
National Science Foundation (CHE-1362006, MRI-1531814)
and the Robert A. Welch Foundation (E-1337).

\providecommand*{\mcitethebibliography}{\thebibliography}
\csname @ifundefined\endcsname{endmcitethebibliography}
{\let\endmcitethebibliography\endthebibliography}{}

\end{document}